\documentclass[12pt,article,nofootinbib]{revtex4}%
\usepackage{amsfonts}
\usepackage{amsmath}
\usepackage{amssymb}
\usepackage{graphicx}
\usepackage{subfig}%
\providecommand{\U}[1]{\protect \rule{.1in}{.1in}}

\renewcommand{\thesection}{\arabic{section}}

\makeatletter \@addtoreset{equation}{section}
\renewcommand{\theequation}{\thesection.\arabic{equation}}
\begin{document}
\preprint{ }
\title{\rightline{\mbox{\small {LPHE-MS-18-03}} \vspace
{1cm}} \textbf{Gapped gravitinos, isospin }$\frac{\mathbf{1}}{2}%
$\textbf{\ particles and }$\mathcal{N}=2$\textbf{ partial breaking}}
\author{E.H Saidi\thanks{E-mail: h-saidi@fsr.ac.ma}}
\affiliation{{\small 1. LPHE-Modeling and Simulations, Faculty of \ Sciences,}}
\affiliation{{\small Mohammed V University, Rabat, Morocco}}
\affiliation{{\small 2. Center of Physics and Mathematics, CPM- Morocco}}
\keywords{}
\pacs{}

\begin{abstract}
Using results on topological band theory of phases of matter and discrete
symmetries, we study topological properties of band structure of physical
systems involving spin $\frac{1}{2}$ and $\frac{3}{2}$ fermions. We apply this
approach to study partial breaking in 4D $\mathcal{N}=2$ gauged supergravity
in rigid limit and we describe the fermionic gapless mode in terms of chiral
anomaly. We study as well the homologue of the usual spin-orbit coupling
$\vec{L}.\vec{S}$, that opens the vanishing band gap for free $s=\frac{1}{2}$
fermions; and show that is precisely given by the central extension of the
$\mathcal{N}=2$ supercurrent algebra in 4D spacetime. We also give comments on
the rigid limit of Andrianopoli et \underline{al} obtained in
\textrm{\cite{12} }and propose an interpretation of energy bands in terms of a
chiral gapless isospin $\frac{1}{2}$-particle (iso-particle). Other features,
such as discrete T- symmetry in FI coupling space, effect of quantum
fluctuations and the link with Nielson-Ninomiya theorem, are also
studied.\newline \textbf{Key words}{\small : }\emph{Topological band theory,
}$\mathcal{N}=2$ \emph{gauged supergravity}, \emph{gapped gravitinos, partial
breaking, chiral anomaly.}

\end{abstract}
\email{E-mail: h-saidi@fsr.ac.ma}
\volumeyear{ }
\volumenumber{ }
\issuenumber{ }
\eid{ }
\maketitle


\section{Introduction}

In the few past years, there has been an intensive interest into topological
band theory in the Brillouin Zone and in 3D effective Chern- Simons field
theories in connection with the phases of matter
\textrm{\cite{1,2,3,31,310,32,33,330,34}}. This interest has been mainly
concerning spin $s=\frac{1}{2}$ topological matter and spin $1$ topological
gauge theories in lower spacetime dimensions; this is because of their
particular properties in dealing with condensed matter systems like
topological insulators and superconductors; and also for the role they play in
quantum Hall effect as well as in the study of boundary states
and\textrm{\ anomalies \cite{4,5,51,52,53,54,55,6}. }By looking to extend some
special features of these studies to systems with spacetime spins beyond
$\frac{1}{2},1$, we fall into supergravity like models where fermionic modes
of higher spins such as relativistic spin $\frac{3}{2}$ are also known to play
a basic role. In this paper, we would like to explore some topological aspects
of band theory of systems having spins less or equal to $2$; and look for a
physical model where topological properties obtained for spins $s=\frac{1}%
{2},1$ can be extended to higher spins. A priori, physical systems with spins
$s\leq2$ may exist in spacetime dimensions $D=d+1\geq3$ where spin $\frac
{3}{2}$ and $2$ particle fields have non trivial gauge degrees of freedom; and
to get started it is then natural to begin by fixing the full spins content of
the physical system we are interested in here; and also define the hamiltonian
model or the field equations describing the full dynamics. To find a physical
system with higher spins where such kinds of studies may be relevant to; and
also identify the appropriate approach that we can use as the starting point,
we give here below two motivations: \textrm{The first one concerns the choice
of a particular system having fermions with different spins, say two types of
fermionic spins }$s=\frac{1}{2}$\textrm{\ and }$\frac{3}{2}$; and the other
regards the tools to use for approaching their band structure. First, by
studying the constructions of \textrm{\cite{6,7,8}}, one comes out with the
conclusion that several topological condensed matter statements based on spin
$\frac{1}{2}$ fermions may be approached by starting with Dirac equation of
$\left(  1+d\right)  $ relativistic theory. From this theory, one may engineer
effective hamiltonians breaking explicitly the SO$\left(  1,d\right)  $
Lorentz symmetry by allowing non linear dispersion relations due to underlying
lattice geometries and interactions. It follows from this description that
quantities like fermionic gapless/gapped modes, chiral ones and edge states
have interpretations in terms massless/massive states, quasi particles with
exotic statistics and anomalies whose explanation requires the use of
topological notions such as manifold boundaries, left/right windings, Berry
connection and Nielson-Ninomiya theorem. To look for extending results on spin
$\frac{1}{2}$ topological matter to spin $\frac{3}{2}$ gravitinos, we then
have to go beyond Dirac equation; for instance by considering the
Rarita-Schwinger equation of gravitinos and try to mimic the analysis done for
spin $\frac{1}{2}$. Even though this is an interesting direction to take
\cite{80,81}, we will not follow this path here because of the complicated
$\mathcal{N}=2$ supergravity interactions making the field equations difficult
to manage. Instead, we will rather use related equations given by extended
supergravity Ward identities \textrm{\cite{9,90}}. The use of these Ward
identities has been motivated by the question to what kind of physical systems
the specific properties of the gravitino\ band structure may serve to.
Recalling the role played by gravitinos in the spontaneous breaking of local
supersymmetry, we immediately come to the point that gapless and gapped
gravitinos can be applied to study the problem of partial supersymmetry
breaking of $\mathcal{N}$- extended vector-like theories. Indeed, in the
example of the effective $\mathcal{N}=2$ gauged supergravity in 4D spacetime,
one has, in addition to bosons (with $s=0,1,2$) and spin $\frac{1}{2}$
fermions, two gravitinos $\left(  \psi_{\alpha \mu}^{1},\psi_{\alpha \mu}%
^{2}\right)  $ forming an isospin $\frac{1}{2}$ particle; that is to say a
doublet under the SU$\left(  2\right)  $ R-symmetry involving pairs of gapless
gravitino modes. The breaking $\mathcal{N}=2\rightarrow \mathcal{N}=1$ requires
then a partial lifting of the degeneracies of mode doublets, which, like in
the case of condensed matter with spin $\frac{1}{2}$ fermions, may be achieved
by turning on a kind of spin-orbit like coupling $\vec{L}.\vec{S}$
\textrm{\cite{10}}. The spin $\frac{3}{2}$ matter study offers therefore a
good opportunity to identify what is the iso-particle hamiltonian including
the homologue of $\vec{L}.\vec{S}$ that induces partial breaking of
supersymmetry. This coupling will be denoted like $\vec{\xi}.\mathcal{\vec{I}%
}$ where $\vec{\xi}$ plays the role of angular momentum $\vec{L}$ and the
isospin $\mathcal{\vec{I}}$ the role of the spin $\vec{S}$. In this regards,
it interesting to recall that spontaneous partial breaking in $\mathcal{N}=2$
supergravity may be done by superHiggs mechanism; which, in $\mathcal{N}=1$
supermultiplet language, a massive $\mathcal{N}=1$ gravitino multiplet can be
created by merging three multiplets: a massless $\mathcal{N}=1$ gravitino
eating a massless $\mathcal{N}=1$ U$\left(  1\right)  $ multiplet and a
$\mathcal{N}=1$ chiral multiplet \textrm{\cite{11}}. But here, the partial
breaking will be done by the isospin-orbit coupling that opens the gap energy
between the two gravitinos. In this study, we will show that the $\vec{\xi
}.\mathcal{\vec{I}}$ coupling is precisely given by the central anomaly of the
$\mathcal{N}=2$ supercurrent algebra in 4D spacetime \textrm{\cite{17,181}%
}.\newline The main purpose of this work is then to use results on topological
band theory of fermionic matter and chiral anomalies as well discrete
symmetries to study partial breaking in $\mathcal{N}=2$ gauged supergravity in
4D. The spacetime fields of our system are given by the fields content of the
standard $\mathcal{N}=2$ supermultiplets; in particular the fields content of
the gravity multiplet, n$_{V}$ vector multiplets and n$_{H}$ matter
multiplets. To perform this study, we will use $\mathcal{N}=2$ supergravity
Ward identities in rigid limit as considered in \textrm{\cite{12}; }and also
study the partial breakings by using the topological approach along the
Nielson-Ninomiya theorem and chiral anomaly. We study as well the effect of
quantum harmonic fluctuations in the FI coupling space; and show that the
result of \textrm{\cite{12} }is not affected by quantum corrections provided a
saturated condition holds. \newline The presentation is as follows: In section
2, we describe some tools on partial breaking in the rigid limit of
$\mathcal{N}=2$ supergravity theory and present the basic equations to start
with. We also give some useful comments. In section 3, we derive the free
hamiltonian of the iso-particles in $\mathcal{N}=2$ gauged supergravity; work
out the isospin-orbit coupling that opens the zero gap between the two
gravitino zero modes and show how time reversing symmetry $\boldsymbol{T}$;
and $\boldsymbol{PT}$ (combined $\boldsymbol{T}$ and parity $\boldsymbol{P}$)
can be implemented. In section 4, we study gapless and gapped gravitinos in
$\mathcal{N}=2$ gauged supergravity, describe the properties of partial
supersymmetry breakings and their interpretation from the view of
Nielson-Ninomiya theorem and chiral anomaly. We discuss also the effect of
quantum fluctuations on partial breaking of $\mathcal{N}=2$ supersymmetry.
Section 5 is devoted to conclusion and comments.

\section{Rigid limit of $\mathcal{N}=2$ Ward identity: case $U\left(
1\right)  $ model}

Following \textrm{\cite{12}}, partial breaking of rigid and local extended
supersymmetries is highly constrained; it can occur in a certain class of
supersymmetric field theories provided one evades some no-go theorems
\textrm{\cite{13,14,15,16}; see also \cite{161,162,163,164,165,166}}. In
global 4D $\mathcal{N}=2$ theories, this was first noticed in
\textrm{\cite{17,18}}; and was explicitly realized in \textrm{\cite{19,20}}
for a model of a self- interacting $\mathcal{N}=2$ vector multiplet in the
presence of $\mathcal{N}=2$ electric and magnetic Fayet-Iliopoulos (FI) terms.
There, it has been shown explicitly that the presence of electric $\vec{\nu}$
and magnetic $\vec{m}$ FI couplings is crucial to achieve partial breaking.
The general conditions for $\mathcal{N}=2$ partial supersymmetry breaking have
been recently elucidated by L. Andrianopoli et \underline{al} in
\textrm{\cite{12} where it has been also shown that} $\vec{\nu}$ and $\vec{m}$
should be non aligned ($\vec{\nu}\wedge \vec{m}\neq \vec{0}$). Their starting
point for deriving the general conditions for partial supersymmetry breaking
in rigid limit\textrm{\footnote{\ Rigid limit is implemented through a
rescaling of the fields contents of the theory and the space time
supercoordinates by using a dimensionless parameter as $\mu=\frac{\Lambda
}{M_{pl}}$. For explicit details see \textrm{\cite{20,13}}.}} was the reduced
$\mathcal{N}=2$\ gauged supergravity Ward identity%
\begin{equation}
\mathcal{V}\delta_{B}^{A}+C_{B}^{A}=\sum_{i=1}^{n_{V}}\delta_{B}\lambda
^{iC}\delta^{A}\lambda_{iC} \label{9b}%
\end{equation}
where the spin $\frac{1}{2}$ fermions $\lambda^{iA}$ and $\lambda_{iB}:=$
$\varepsilon_{BA}g_{i\bar{j}}\lambda^{\bar{j}A}$ refer to the chiral and
antichiral projections of the gauginos respectively. Here, the SO$\left(
1,3\right)  $ spacetime spin index of the $\lambda^{iA}$ fermions has been
omitted for simplicity; while we have exhibited the two other indices $A$
and\textrm{ }$i.$ The $A=1,2$ refers to the isospin $\frac{1}{2}$
representation of the SU$\left(  2\right)  _{R}$ symmetry of the
$\mathcal{N}=2$ supersymmetric algebra seen that $\mathcal{N}=2$ gauginos are
isodoublets under SU$\left(  2\right)  _{R}$; this index is lowered and rised
by the antisymmetric tensor $\varepsilon^{AB}$ and its inverse $\varepsilon
_{BA}$. The index $i=1,...,n_{V}$ designates the number of $\mathcal{N}=2$
vector multiplets in the Coulomb branch of the $\mathcal{N}=2$ \ gauged
supergravity theory. Notice also that the quantity $\left(  \delta_{B}%
\lambda^{iA}\right)  $ is a convention notation for the $\mathcal{N}=2$
supersymmetric transformation of gauginos which is given by $\delta
_{susy}\lambda^{iA}=\left(  \delta_{B}\lambda^{iA}\right)  \epsilon^{B}$ with
the two fermions $\epsilon^{A}=\left(  \epsilon^{1},\epsilon^{2}\right)  $
standing for the supersymmetric transformation parameters. \textrm{In the
relation (\ref{9b})}, the right hand side is restricted to the pure Coulomb
branch and so corresponds to the rigid limit of the following local
identities
\begin{equation}
\sum_{i}\alpha_{i}\delta_{B}\lambda^{iC}\delta^{A}\lambda_{iC}=\mathcal{\tilde
{V}}\delta_{B}^{A}-\sum_{u}\alpha_{u}\delta_{B}\zeta^{u}\delta^{A}\zeta
_{u}-\sum_{\mu,\nu}\alpha_{0}\delta^{A}\psi_{\nu C}\Gamma^{\mu \nu}\delta
_{B}\psi_{\mu}^{C} \label{re1}%
\end{equation}
The left hand side of (\ref{9b}) contains two basic terms namely the rigid
limit of the scalar potential $\mathcal{V}\delta_{B}^{A}$ and an extra
traceless constant matrix,
\begin{equation}
C_{B}^{A}=\vec{\xi}.\left(  \vec{\tau}\right)  _{B}^{A}\qquad,\qquad TrC=0
\label{ca}%
\end{equation}
This hermitian traceless matrix can be interpreted as an anomalous central
extension in the $\mathcal{N}=2$ supersymmetric current algebra
\textrm{\cite{18,13,181}}; it only affects the commutator of two supersymmetry
transformations of the gauge field \textrm{\cite{13,20} and contains data on
hidden gravity and matter sectors}. Recall that the basic anticommutator of
the $\mathcal{N}=2$ supercurrent algebra is,
\begin{equation}
\left \{  \mathcal{J}^{0A}\left(  x\right)  ,\int d^{3}y\mathcal{\bar{J}}%
_{B}^{0}\left(  y\right)  \right \}  =\delta_{B}^{A}\sigma_{\mu}\mathcal{T}%
^{\mu0}+C_{B}^{A} \label{ac}%
\end{equation}
where $\mathcal{J}_{\alpha A}^{0}\left(  x\right)  ,$ $\mathcal{\bar{J}}%
_{\dot{\alpha}A}^{0}\left(  x\right)  $ and $\mathcal{T}_{\mu}^{0}\left(
x\right)  $ are the time components of the supersymmetric current densities
$\mathcal{J}_{\alpha A}^{\nu},$ $\mathcal{\bar{J}}_{\dot{\alpha}A}^{\nu}$ and
$\mathcal{T}_{\mu}^{\nu}$. The time component densities in the current
superalgebra (\ref{ac}) are related to $Q_{\alpha}^{A}$, $\bar{Q}_{\dot
{\alpha}B}$ and $P_{\mu}$ charges of the $\mathcal{N}=2$ supersymmetric
QFT$_{4}$ in the usual manner; for example
\begin{equation}
Q_{\alpha A}=\int d^{3}x\mathcal{J}_{\alpha A}^{0}\qquad,\qquad P_{\mu}=\int
d^{3}x\mathcal{T}_{\mu}^{0}%
\end{equation}
obeying $Q_{B}\bar{Q}^{A}+\bar{Q}^{A}Q_{B}\sim \delta_{B}^{A}\sigma^{\mu}%
P_{\mu}$; the usual globally defined $\mathcal{N}=2$ supersymmetric algebra
with $C_{B}^{A}$ constrained to vanish.\newline By comparing eq(\ref{9b}) with
the general form of the Ward identities eq(\ref{re1}), we deduce that the
$C_{B}^{A}$ term captures the contribution of the fermion shifts to the Ward
identity coming from the rigid limit of the hidden gravity ($\delta^{A}%
\psi_{\nu C}\Gamma^{\mu \nu}\delta_{B}\psi_{\mu}^{C}$) and the matter
($\delta_{B}\zeta^{u}\delta^{A}\zeta_{u}$) branches. For the simple example of
an abelian $U\left(  1\right)  $ gauge multiplet ($n_{V}=1$), the anomaly
isovector $\vec{\xi}=Tr\left(  \vec{\tau}C\right)  $ has been realised in
terms of the electric $\vec{\nu}$ and the magnetic $\vec{m}$ FI coupling
constant isovectors of the Coulomb branch of the effective $\mathcal{N}%
=2$\ U$\left(  1\right)  $ gauge theory as follows\textrm{\footnote{\ The
exact expression found in \cite{12}\ is $\vec{\xi}=2\vec{\nu}\wedge \vec{m}.$
Here the factor $2=(\sqrt{2})^{2}$ has been absorbed by scaling the FI
couplings.}}%
\begin{equation}
\vec{\xi}=\vec{\nu}\wedge \vec{m}\qquad,\qquad \vec{m}\neq \mathbb{R}^{\ast}%
\vec{\nu} \label{10b}%
\end{equation}
obeying the remarkable property $\vec{\xi}.\vec{\nu}=0$ and $\vec{\xi}.\vec
{m}=0$; see (\ref{20}). Moreover, partial breaking of supersymmetry takes
place at \textrm{\cite{12,13},}
\begin{equation}
\mathcal{V}=\left \vert \vec{\xi}\right \vert \geq0 \label{kc}%
\end{equation}
This relation will be used later on when considering topological aspects of
gapless fermions (subsection 4.1) as well as harmonic fluctuations (subsection
4.2) ; but before that let us give other comments regarding (\ref{10b}%
-\ref{kc}). \newline First, notice that in order to have a non zero $\vec{\xi
}$, it is sufficient to take the following particular and simple choice%
\begin{equation}
\vec{\nu}=\left(
\begin{array}
[c]{c}%
\nu_{x}\\
0\\
0
\end{array}
\right)  \qquad,\qquad \vec{m}=\left(
\begin{array}
[c]{c}%
0\\
m_{y}\\
0
\end{array}
\right)  \qquad,\qquad \vec{\xi}=\left(
\begin{array}
[c]{c}%
0\\
0\\
\xi_{z}%
\end{array}
\right)  \label{sh}%
\end{equation}
satisfying $\vec{\nu}.\vec{m}=0$ and $\vec{\nu}\wedge \vec{m}\neq \vec{0}$. This
particular choice shows that a quadratic term of type
\[
\frac{\mathrm{\gamma}_{\bot}}{2}\left \vert \vec{m}\right \vert \times \left \vert
\vec{\nu}\right \vert
\]
like the one appearing in eq(\ref{v}), comes necessary for the contribution of
the $\vec{\xi}$- direction normal to the $\left(  \vec{\nu},\vec{m}\right)  $
plane; that is to say:
\begin{equation}
\vec{\xi}=\vec{0}\qquad \Rightarrow \qquad \mathrm{\gamma}_{\bot}=0
\end{equation}
This implication is obviously not usually true since for non orthogonal
$\vec{\nu}$ and $\vec{m}$, we have $\vec{\nu}.\vec{m}=\left \vert \vec
{m}\right \vert \times \left \vert \vec{\nu}\right \vert \cos \theta \neq0$ as far
as $\theta \neq \pm \frac{\pi}{2}$ $\operatorname{mod}2\pi$. The trick
$\theta=\pm \frac{\pi}{2}$ will help us to detect the effect of $\vec{\xi}$
especially when studying quantum fluctuations around the $\mathcal{N}=2$
supersymmetric ground states $\mathcal{V}=0$ and $\mathcal{V}=\left \vert
\vec{\xi}\right \vert $. \newline Second, observe that by setting
$\tau_{\left[  kl\right]  }=\frac{1}{2}\varepsilon_{kln}\tau^{n}$,
$\xi^{\left[  lk\right]  }=\varepsilon^{klm}\xi_{n}$ and $\varepsilon
_{kln}\varepsilon^{klm}=2\delta_{n}^{m}$, it follows that $\xi^{\left[
kl\right]  }\tau_{\left[  kl\right]  }=\xi_{i}\tau^{i}$; and then the central
extension matrix (\ref{ca}) can be also expressed like
\begin{equation}
C_{B}^{A}=\xi^{\left[  kl\right]  }\left(  \tau_{\left[  kl\right]  }\right)
_{B}^{A}\qquad,\qquad \tau_{\left[  kl\right]  }=\frac{1}{4i}\left[  \tau
_{k},\tau_{l}\right]
\end{equation}
This way of expressing $C_{B}^{A}$ is interesting since, supported by the
dimensional argument, it gives an idea on how to realize the factor
$\xi^{\left[  kl\right]  }$ in terms of the electrical $\nu^{k}$ and magnetic
$m^{l}$ couplings of FI. Antisymmetry implies the natural factorisation
\begin{equation}
\xi^{\left[  kl\right]  }=\nu^{k}m^{l}-\nu^{l}m^{k}\qquad,\qquad \xi^{\left[
xy\right]  }=\nu^{x}m^{y}-\nu^{y}m^{x} \label{10c}%
\end{equation}
which is nothing but the Andrianopoli et \underline{al} factorisation
(\ref{10b}). We expect that this trick can also help to find extension the
$\mathcal{N}=2$ realisation (\ref{10b}) to higher supergravities; in
particular to $\mathcal{N}=4$ theory in rigid limit where there is no matter
branch; this generalisation will not be considered here. Notice that for the
simple choice (\ref{sh}), we have the diagonal matrix,%
\begin{equation}
C_{B}^{A}=\left(
\begin{array}
[c]{cc}%
\nu_{x}m_{y} & 0\\
0 & -\nu_{x}m_{y}%
\end{array}
\right)
\end{equation}
showing that in the rest frame, we have $\delta_{B}^{A}\sigma_{\mu}%
\mathcal{T}^{\mu0}=\delta_{B}^{A}\mathcal{V}$ andd then the $\mathcal{N}=2$
current algebra (\ref{ac}) splits into two $\mathcal{N}=1$ copies with right
hand energy densities given by $\mathcal{V}\pm \nu_{x}m_{y}$.\newline Third, a
non vanishing $\vec{\xi}$ requires in general non collinear $\vec{\nu}$ and
$\vec{m}$ vectors; and so the unit vectors $\vec{e}_{\nu}=\frac{\vec{\nu}%
}{\left \vert \vec{\nu}\right \vert }$, $\vec{e}_{m}=\frac{\vec{m}}{\left \vert
\vec{m}\right \vert }$ generate a 2-dimensional plane with a normal vector
given by $\vec{e}_{\xi}=\frac{\vec{\xi}}{\left \vert \vec{\xi}\right \vert }$.
These three vectors form altogether a 3D vector basis of $\mathbb{\tilde{R}%
}^{3}$ that we terme as the 3D iso-space,
\begin{equation}
\vec{e}_{\nu}\text{ };\text{ }\vec{e}_{m}\text{ };\text{ }\vec{e}_{\xi}%
\qquad,\qquad \vec{e}_{\xi}=\vec{e}_{\nu}\wedge \vec{e}_{m} \label{20}%
\end{equation}
By the terminology 3D iso-space, we intend to use its similarity with the
usual Euclidean space $\mathbb{R}^{3}$ of classical mechanics of point like
particles to propose a physical interpretation of (\ref{10b}) by using the
notion of isospin $I=\frac{1}{2}$ particle; this will be done in section
3.\newline Notice moreover the following features:

\begin{itemize}
\item the relation (\ref{10b}) concerns an effective $\mathcal{N}=2$ U$\left(
1\right)  $ gauge theory with one gauge supermultiplet $n_{V}=1$. The general
expression of $\vec{\xi}$, extending (\ref{10b}), as well as the general form
of the scalar potential energy $\mathcal{V}$ associated with generic U$\left(
1\right)  ^{n_{V}}$ effective gauge theories reduced to FI couplings, have
been shown to be functions of characteristic data of the special geometry of
the scalar manifold. They are given by the following factorisations%
\begin{align}
\mathcal{V}  &  =\frac{1}{2}\delta_{ab}\mathcal{P}^{aM}\mathcal{S}%
_{MN}\mathcal{P}^{bN}\label{ae}\\
\xi_{a}  &  =\frac{1}{2}\varepsilon_{abc}\mathcal{P}^{bM}\mathcal{C}%
_{MN}\mathcal{P}^{cN} \label{ce}%
\end{align}
where $\mathcal{P}^{aM}=\left(  m^{aI},\nu_{I}^{a}\right)  ^{t}$ are moment
maps carrying quantum numbers of SU$\left(  2\right)  _{R}\times SP\left(
2n_{V},R\right)  $; $\mathcal{C}_{MN}$ the metric of $SP\left(  2n_{V}%
,R\right)  $; and $\mathcal{S}_{MN}$ is a symmetric matrix of the form%
\begin{equation}
\mathcal{S}_{MN}=\left(
\begin{array}
[c]{cc}%
\mathcal{I+RI}^{-1}\mathcal{R} & -\mathcal{RI}^{-1}\\
-\mathcal{I}^{-1}\mathcal{R} & \mathcal{I}^{-1}%
\end{array}
\right)  \label{sn}%
\end{equation}
encoding data on the scalar manifold of the $\mathcal{N}=2$ theory, see
\textrm{\cite{12,13} for more details}. For the example of an abelian
$U\left(  1\right)  $ gauge model, the $\xi_{a}$ is as in eq(\ref{10b}) while
the scalar potential (\ref{ae}) has also the following remarkable quadratic
shape
\begin{equation}
\mathcal{V}=\mathrm{\alpha}\left \vert \vec{m}\right \vert ^{2}+\mathrm{\beta
}\left \vert \vec{\nu}\right \vert ^{2}+\frac{\mathrm{\gamma}}{2}\left \vert
\vec{m}\right \vert \times \left \vert \vec{\nu}\right \vert \label{v}%
\end{equation}
with $4\alpha \beta>\gamma^{2}>0$ and $\alpha,\beta$\ assumed positive for
later use. Notice that here $\mathrm{\gamma}$ should\textrm{\ be viewed as the
su}m $\mathrm{\gamma}_{\Vert}+\mathrm{\gamma}_{\bot}$ with $\mathrm{\gamma
}_{\Vert}$ describing the coupling in the $\left(  \vec{m},\vec{\nu}\right)  $
plane and $\mathrm{\gamma}_{\bot}$ in the normal $\vec{m}\wedge \vec{\nu}$
directions; see also eqs(\ref{xa}-\ref{ax}).

\item By substituting (\ref{sn}) and $\mathcal{P}^{a}=\left(  m^{a},\nu
^{a}\right)  $ into (\ref{ae}), we learn that the real parameters
$\mathrm{\alpha}$, $\mathrm{\beta}$ and $\mathrm{\gamma}$ in above scalar
potential have indeed a geometric interpretation in terms of the effective
prepotential $\mathcal{F}$ of the $\mathcal{N}=2$ special geometry. For
example the parameters $\frac{\mathrm{\gamma}}{2}$ in (\ref{v}) depends both
on the real $\mathcal{R}$ and imaginary $\mathcal{I}$ parts of the second
derivative of $\mathcal{F}$.

\item The scalar potential (\ref{v}) has a particular dependence on
$\left \vert \vec{m}\right \vert $ and $\left \vert \vec{\nu}\right \vert $; it
can be presented as quadratic form $\mathcal{V}=P^{i}G_{ij}P^{j}$ with $P^{i}$
and metric $G_{ij}$ as follows
\begin{equation}
\mathcal{V}=\left(  \left \vert \vec{m}\right \vert ,\left \vert \vec{\nu
}\right \vert \right)  \left(
\begin{array}
[c]{cc}%
\mathrm{\alpha} & \frac{\mathrm{\gamma}}{2}\\
\frac{\mathrm{\gamma}}{2} & \mathrm{\beta}%
\end{array}
\right)  \left(
\begin{array}
[c]{c}%
\left \vert \vec{m}\right \vert \\
\left \vert \vec{\nu}\right \vert
\end{array}
\right)  \label{w}%
\end{equation}
with $\det G=\alpha \beta-\frac{\gamma^{2}}{4}$. This form will diagonalised
later on for explicit calculations.

\item The above $\mathcal{V}$ is might be viewed as a special potential; a
more general expression would involve more free parameters as shown here
below,%
\begin{align}
\mathcal{V}  &  =V_{0}+\varrho_{a}\nu^{a}+w_{a}m^{a}+B_{ab}\nu^{a}%
m^{b}+\nonumber \\
&  A_{ab}\nu^{a}\nu^{b}+C_{ab}m^{a}m^{b} \label{xe}%
\end{align}
where $V_{0}$ is a number that depends on the VEVs of the scalar fields and
the parameters of the effective $\mathcal{N}=2$ theory like masses and gauge
coupling constants. The $\varrho_{a}$, $w_{a}$ are two isovectors scaling in
same manner as the FI constants; and $A_{ab}$, $B_{ab}$, $C_{ab}$ are
dimensionless real 3$\times$3 matrices; $A_{ab}$ and $C_{ab}$ are symmetric;
but $B_{ab}$ is a general matrix. These moduli may characterise as well the
scalar manifold of the effective $\mathcal{N}=2$ supergravity and likely
external fields as suspected from table (\ref{13}); see also eq(\ref{S}) where
$\vec{w}$ of the $w_{a}m^{a}$ is interpreted in terms of an external
iso-magnetic field.
\end{itemize}

\  \  \newline Finally, notice that by giving these somehow explicit details on
the $n_{V}=1$ theory, we intend to use its simple properties to derive the
iso-particle proposal and build the isospin-orbit coupling in $\mathcal{N}=2$
supergravity mentioned in the introduction. We will also use these tools to
study the isospin $\frac{1}{2}$ particle as well as hidden discrete symmetries
that capture data on the topological phases of the right hand of the
$\mathcal{N}=2$ supersymmetry current algebra (\ref{ac}).

\section{Isospin $\frac{1}{2}$ particle proposal}

\begin{figure}[ptbh]
\begin{center}
\includegraphics[scale=0.8]{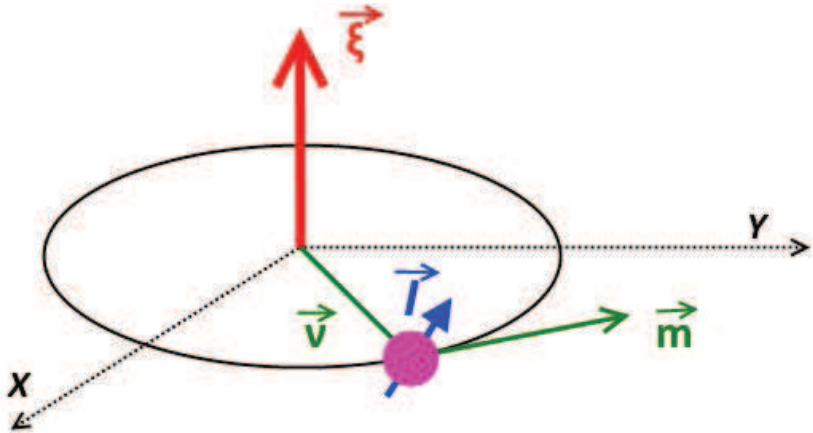}
\end{center}
\par
\vspace{-0.5 cm}\caption{ \textit{A classical quasi-particle with angular
momentum }$\vec{\xi}=\vec{\nu}\wedge \vec{m}$\textit{ in FI coupling space
parameters. The electric FI coupling }$\vec{\nu}$\textit{ is viewed as
position vector }$\vec{r}$\textit{ and the magnetic coupling }$\vec{m}%
$\textit{ as momentum }$\vec{p}$\textit{. In addition to }$\left(  \vec{\nu
},\vec{m}\right)  $\textit{, the quasi-particle carries also an intrinsic
isospin charge }$I=\frac{1}{2}$\textit{ as well as unit }$U\left(  1\right)
_{elec}\times U\left(  1\right)  _{mag}$\textit{ charge due to the gauging of
abelian isometries of }$N=2$\textit{ gauged supergravity. This image may be
put in correspondence with an electron spining around a nucleus.}}%
\label{1}%
\end{figure}

The Andrianopoli et \underline{al} realisation (\ref{10b}) of the rigid
anomaly isovector $\vec{\xi}=\vec{\nu}\wedge \vec{m}$ in effective
$\mathcal{N}=2$ supersymmetric gauge theory is interesting and is very
suggestive; see figure (\ref{1}) for illustration. This is because of the
wedge product $\vec{\nu}\wedge \vec{m}$ that allows us to establish a
correspondence between properties of partial supersymmetry breaking and the
electronic band theory with $\Delta_{soc}\vec{L}.\vec{S}$ spin-orbit coupling
turned on ($\Delta_{soc}\neq0$).\newline Indeed, the axial vector $\vec{\xi}=\vec{\nu}\wedge \vec{m}$, to
which we refer below to as \emph{Andrianopoli et }\underline{al} orbital
vector, has the same form of the usual angular momentum vector,
\begin{equation}
\vec{L}=\vec{r}\wedge \vec{p} \label{am}%
\end{equation}
of a 3D classical particle with coordinate position $\vec{r}$ and momentum
$\vec{p}$. By comparing the $\vec{\nu}\wedge \vec{m}$ formula of (\ref{10b})
with the above $\vec{r}\wedge \vec{p}$, it follows that the FI electric
coupling $\vec{\nu}$ may be put in correspondence with the vector $\vec{r}$;
and the magnetic $\vec{m}$ with the vector $\vec{p}$. Hence, we have the
following schematic picture linking the physics of classical particles
(electron) to the physics of iso-particle of $\mathcal{N}=2$ gauged
supergravity (gravitinos and gauginos),
\begin{equation}%
\begin{tabular}
[c]{lll}%
\  \ vectors in $\mathbb{R}^{3}:$ electron & $\qquad \leftrightarrow \qquad$ &
\  \ iso-vectors in $\mathbb{\tilde{R}}^{3}:$ gravitinos\\ \hline \hline
\  \  \ particle & $\qquad$\ $:\qquad$ & \  \ iso-particle\\
$\  \  \  \  \  \left(  \vec{r};\vec{p}\right)  $ & $\qquad$\ $:\qquad$ &
$\  \  \  \  \left(  \vec{\nu};\vec{m}\right)  $\\
$\  \  \  \left(  \vec{r}+\delta \vec{r};\vec{p}+\delta \vec{p}\right)  $ &
$\qquad$\ $:\qquad$ & $\  \  \  \  \left(  \vec{\nu}+\delta \vec{\nu};\vec
{m}+\delta \vec{m}\right)  $\\
isotropy SO$\left(  3\right)  $ & $\qquad$\ $:\qquad$ & R-symmetry SU$\left(
2\right)  $\\
orbital moment $\vec{L}=\vec{r}\wedge \vec{p}$ & $\qquad$\ $:\qquad$ & orbital
moment $\vec{\xi}=\vec{\nu}\wedge \vec{m}$\\
hamiltonian h$\left(  \vec{r};\vec{p}\right)  $ & $\qquad$\ $:\qquad$ &
hamiltonian h$\left(  \vec{\nu};\vec{m}\right)  $\\
\  \  \ spin \ $\vec{S}$ & $\qquad$\ $:\qquad$ & \  \ isospin \ $\mathcal{\vec
{I}}$\\
gauge summetry $U\left(  1\right)  _{em}$ & $\qquad$\ $:\qquad$ & gauge
summetry $U\left(  1\right)  _{elec}\times U\left(  1\right)  _{mag}%
$\\ \hline \hline
\end{tabular}
\  \  \  \label{13}%
\end{equation}%
\[
\]
On the left hand side of this table, the Euclidian $\mathbb{R}^{3}$ space is
the usual 3d- space with SO$\left(  3\right)  $ isotropy symmetry. In this
real space lives bosons and fermions; in particular fermions with intrinsic
properties like spin $\frac{1}{2}$ particles with symmetry
\begin{equation}
SU\left(  2\right)  _{spin}\sim SO\left(  3\right)  \label{r}%
\end{equation}
On right hand side, the $\mathbb{\tilde{R}}^{3}$ is an iso-space with isotropy
symmetry $SO\left(  3\right)  _{R}$ given by the R-symmetry $SU\left(
2\right)  _{R}$ of the $\mathcal{N}=2$ supersymmetric algebra. This is a
global symmetry group that will be imagined here as a global isospin group
$SU\left(  2\right)  _{isospin}$ characterising the quasi-particle of figure
(\ref{1}). Thus, the homologue of the real space symmetry (\ref{r}) is given
by,%
\begin{equation}
SU\left(  2\right)  _{isospin}\sim SU\left(  2\right)  _{R}\sim SO\left(
3\right)  _{R}%
\end{equation}
Matter in the iso-space $\mathbb{\tilde{R}}^{3}$ is then given by
quasi-particles carrying isospin charges under $SU\left(  2\right)  _{R}$; in
particular the isospin $I=\frac{1}{2}$ describing the two gravitinos and the
$n_{V}$ pairs of gauginos of the Coulomb branch of the $\mathcal{N}=2$ gauged
supergravity. Recall that in this theory, the particle content belongs to
three $\mathcal{N}=2$ supermultiplets namely the gravity $\boldsymbol{G}%
_{\mathcal{N}=2}$, the vector $\boldsymbol{V}_{\mathcal{N}=2}$ and the matter
$\boldsymbol{H}_{\mathcal{N}=2}$. The two first ones are recalled here after%
\[
\]%
\begin{equation}%
\begin{tabular}
[c]{|l|l|l|l|}\hline
$\  \  \mathcal{N}=2$ multiplets \  \  & \  \  \  \ field content in spin &
$\  \  \  \  \ $spin s $\  \  \  \  \ $ & $\  \ $isospin I$\  \  \  \  \ $\\ \hline
\  \  \ gravity $\boldsymbol{G}_{\mathcal{N}=2}$ & $\  \left.
\begin{array}
[c]{ccc}%
\text{{\small graviton}} & : & 2\\
\text{{\small gravitinos \ }}\psi^{A}\text{ \  \ } & : & 2\times \frac{3}{2}\\
\text{{\small graviphoton }}A_{\mu}^{1} & : & 1
\end{array}
\right.  $ & $\  \  \  \  \  \left.
\begin{array}
[c]{c}%
2\\
\frac{3}{2}\\
1
\end{array}
\right.  $ & $\  \  \  \  \  \left.
\begin{array}
[c]{c}%
0\\
\frac{1}{2}\\
0
\end{array}
\right.  $\\ \hline
\  \  \ vector $\boldsymbol{V}_{\mathcal{N}=2}$ & $\  \left.
\begin{array}
[c]{ccc}%
\text{ \  \ {\small vector \ }}A_{\mu}^{2}\text{{\small \  \  \ }} & : & 1\\
\text{{\small gauginos \ }}\lambda^{A}\text{ \  \  \  \ } & : & 2\times \frac
{1}{2}\\
\text{{\small scalars}} & : & 2\times0
\end{array}
\right.  $ & $\  \  \  \  \  \left.
\begin{array}
[c]{c}%
1\\
\frac{1}{2}\\
0
\end{array}
\right.  $ & $\  \  \  \  \  \left.
\begin{array}
[c]{c}%
0\\
\frac{1}{2}\\
0
\end{array}
\right.  $\\ \hline
\end{tabular}
\  \label{is}%
\end{equation}%
\[
\]
The field content includes the fermions (gravitinos and gauginos) having a non
trivial isospin charge. It contains also two spin $s=1$ gauge fields $A_{\mu
}^{M}$ ( graviphoton A$_{\mu}^{1}$ and Coulomb A$_{\mu}^{2}$) with
\begin{equation}
U\left(  1\right)  _{elec}\times U\left(  1\right)  _{mag} \label{si}%
\end{equation}
gauge transformations given by abelian isometries of the scalar manifold of
the supergravity theory. The fermionic fields $\digamma^{A}=\psi^{A}%
,\lambda^{A}$ carry a unit $U\left(  1\right)  _{elec}\times U\left(
1\right)  _{mag}$ charge; and interact with the gauge vector fields $A_{\mu
}^{M}$ through the minimal coupling $D_{\mu}\digamma^{A}$ where the covariant
derivative $D_{\mu}=\partial_{\mu}+\vartheta_{M}A_{\mu}^{M}$ with
electric/magnetic coupling $\vartheta_{M}$; see
\textrm{\cite{19,20,15,16,22,23,24}} for other features. \newline In
(\ref{13}), we have moreover an exotic variable $\tau$, playing the role of
the real time t of the left hand side of the table. This $\tau$ may be
imagined in terms of energy scale variable; and hence one is left with running
complings $\vec{\nu}=\vec{\nu}\left(  \tau \right)  $ and $\vec{m}=\vec
{m}\left(  \tau \right)  $ with%
\begin{equation}
\vec{m}\left(  \tau \right)  \sim \frac{d\vec{\nu}\left(  \tau \right)  }{d\tau
}\qquad \leftrightarrow \qquad \vec{p}\left(  t\right)  \sim \frac{d\vec{r}\left(
t\right)  }{dt}%
\end{equation}
In what follows, we assume that the classical correspondence (\ref{13}) is
valid as well at quantum level and study the energy band properties of the
isospin $\frac{1}{2}$ particles (gravitinos and gauginos) of the
$\mathcal{N}=2$ gauged supergravity.

\subsection{Deriving the free hamiltonian of iso-particle}

Here, we use (\ref{13}) to build the free hamiltonian $\boldsymbol{h}=h\left(
\nu,m\right)  $ of the iso-particle and study its classical and quantum
behaviours. We also comment on some interacting terms appearing in the scalar
potential (\ref{xe}).

\subsubsection{Classical description}

Using the proposal (\ref{13}), the free hamiltonian $\boldsymbol{h}$ of the
classical iso-particle is given by the scalar potential of the supergravity
theory. It is just the energy density of the supergravity theory,
\begin{equation}
\boldsymbol{h}=\mathcal{V}\left(  \vec{\nu},\vec{m}\right)  \label{ph}%
\end{equation}
Because this energy is quadratic in $\vec{m}$ and $\vec{\nu}$ as shown by the
rigid limit of \cite{12}, the $\boldsymbol{h}$ describes then the free
dynamics of a classical iso-particle in the 6D phase space $\mathbb{\tilde{R}%
}^{3}\times \mathbb{\hat{R}}^{3}$ parameterised by the FI coupling parameters.
By using eqs(\ref{v}-\ref{w}), we have
\begin{equation}
\boldsymbol{h}=\mathrm{\alpha}\left \vert \vec{m}\right \vert ^{2}%
+\mathrm{\beta}\left \vert \vec{\nu}\right \vert ^{2}+\frac{\mathrm{\gamma
}_{\Vert}}{2}\left \vert \vec{m}\right \vert \times \left \vert \vec{\nu
}\right \vert \label{hp}%
\end{equation}
where $\mathrm{\alpha}$, $\mathrm{\beta}$ and the planar $\mathrm{\gamma
}_{\Vert}$ are three real parameters that have an interpretation in the
special geometry of the scalar manifold of the $\mathcal{N}=2$ effective
theory. Here, they will be given an interpretation in terms of an effective
mass $\mathrm{\mu}$ and a frequency\ $\omega$ with relationships an in
(\ref{mk}) and (\ref{4}). Notice the following useful features:

\begin{itemize}
\item The above $\boldsymbol{h}$ has the form of a classical harmonic
oscillator energy $\frac{p_{x}^{2}}{2M}+\frac{Mw^{2}}{2}x^{2}$; so one can
take advantage from this feature to learn more about the properties of the
iso-particle of the $\mathcal{N}=2$ gauged supergravity.

\item The notation $\mathrm{\gamma}_{\Vert}$ in (\ref{hp}) is to distinguish
it from another contribution $\mathrm{\gamma}_{\perp}$ to be turned on later
when switching on $\vec{\xi}.\mathcal{\vec{I}}$. By using the two types of
vector products, a general quadratic term like $\left \vert \vec{m}\right \vert
\times \left \vert \vec{\nu}\right \vert $ has in general the typical form
\begin{equation}
\frac{\mathrm{\gamma}}{2}\left \vert \vec{m}\right \vert \times \left \vert
\vec{\nu}\right \vert =\frac{\Lambda}{2}\vec{m}.\vec{\nu}+\frac{\Lambda
^{\prime}}{2}\left \Vert \vec{m}\wedge \vec{\nu}\right \Vert
\end{equation}
showing that $\frac{\mathrm{\gamma}}{2}$ may come from two sources: $\left(
i\right)  $ from a scalar product like $\frac{\Lambda}{2}\vec{m}.\vec{\nu}$;
and/or $\left(  ii\right)  $ from the norm of the wedge product of the two
vectors as follows%
\begin{align}
\frac{\Lambda}{2}\vec{m}.\vec{\nu}  &  =\frac{\mathrm{\gamma}_{\Vert}}%
{2}\left \vert \vec{m}\right \vert \times \left \vert \vec{\nu}\right \vert
\qquad,\qquad \mathrm{\gamma}_{\Vert}=\Lambda \cos \theta \label{xa}\\
\frac{\Lambda^{\prime}}{2}\left \Vert \vec{m}\wedge \vec{\nu}\right \Vert  &
=\frac{\mathrm{\gamma}_{\perp}}{2}\left \vert \vec{m}\right \vert \times
\left \vert \vec{\nu}\right \vert \qquad,\qquad \mathrm{\gamma}_{\perp}%
=\Lambda^{\prime}\left \vert \sin \theta \right \vert \label{ax}%
\end{align}

\item In order to fix a freedom in the signs of $\mathrm{\alpha}$,
$\mathrm{\beta}$, $\mathrm{\gamma}_{\Vert}$; we assume that the discriminant
of the $G_{ij}$ metric of (\ref{hp}) is positive definite,
\begin{equation}
\det G_{ij}=\mathrm{\alpha \beta-}\frac{\mathrm{\gamma}_{\Vert}^{2}}{4}>0
\label{det}%
\end{equation}
As this discriminant is non sensitive to $\left(  \mathrm{\alpha
},\mathrm{\beta},\mathrm{\gamma}_{\Vert}\right)  \rightarrow \left(
-\mathrm{\alpha},-\mathrm{\beta},-\mathrm{\gamma}_{\Vert}\right)  $; we
restrict $\mathrm{\alpha}$, $\mathrm{\beta}$ to be both positive; this
constraint is also needed for $\boldsymbol{h}$ in order to be bounded from
below; this an important condition for quantisation of fluctuation of couplings.

\item The omission of the zero value in $\det G$ is because for
$\mathrm{\alpha \beta-}\frac{\mathrm{\gamma}_{\Vert}^{2}}{4}=0$, the above
hamiltonian (\ref{hp}) reduces to
\begin{equation}
\boldsymbol{h}_{\eta}=Z_{\eta}^{2}\qquad with\qquad Z_{\pm}^{2}=(\left \vert
\vec{m}\right \vert \sqrt{\mathrm{\alpha}}\pm \left \vert \vec{\nu}\right \vert
\sqrt{\mathrm{\beta}})^{2}%
\end{equation}
ruling out harmonic oscillations needed for quantum fluctuations; see
eq(\ref{4}). Nevertheless, the saturated limit captures as well an interesting
data; it will be discussed in subsection 4.2.
\end{itemize}

With these features in mind, we are now in position to deal with the
hamiltonian (\ref{hp}). To that purpose, we perform a linear change of
variables $\left(  \left \vert \vec{m}\right \vert ,\left \vert \vec{\nu
}\right \vert \right)  \rightarrow \left(  \left \vert \vec{m}^{\prime
}\right \vert ,\left \vert \vec{\nu}^{\prime}\right \vert \right)  $ in order to
put $\boldsymbol{h}$ into a normal form as follows
\begin{equation}
\boldsymbol{h}_{0}=\frac{1}{2\mathrm{\mu}}\sum_{a=1}^{3}\left(  m_{a}^{\prime
}\right)  ^{2}+\frac{\mathrm{\kappa}}{2}\sum_{a=1}^{3}\left(  \nu_{a}^{\prime
}\right)  ^{2} \label{h}%
\end{equation}
where now $\frac{1}{2\mathrm{\mu}}\vec{m}^{\prime2}$ stands for "kinetic
energy" and $\frac{\mathrm{\kappa}}{2}\vec{\nu}^{\prime2}$ for the "potential
energy". The new $\left \vert \vec{m}^{\prime}\right \vert ,$ $\left \vert
\vec{\nu}^{\prime}\right \vert $ are related to the old $\left \vert \vec
{m}\right \vert ,$ $\left \vert \vec{\nu}\right \vert $ ones by
\begin{equation}
\left \vert \vec{m}^{\prime}\right \vert =a\left \vert \vec{\nu}\right \vert
+b\left \vert \vec{m}\right \vert \qquad,\qquad \left \vert \vec{\nu}^{\prime
}\right \vert =c\left \vert \vec{\nu}\right \vert +d\left \vert \vec
{m}\right \vert
\end{equation}
with $ad-bc=1$, that diagonalise the metric in (\ref{w}). The resulting
positive mass $\mathrm{\mu}$ and $\mathrm{\kappa}=\mathrm{\mu}\omega^{2}$
(oscillation frequency) are functions of the $\mathrm{\alpha}$, $\mathrm{\beta
}$, $\mathrm{\gamma}_{\Vert}$ parameters; their explicit expressions read as
follows
\begin{equation}
\frac{1}{\mathrm{\mu}}=\alpha+\beta+\sqrt{\left(  \alpha-\beta \right)
^{2}+\mathrm{\gamma}_{\Vert}^{2}}\qquad,\qquad \mathrm{\kappa}=\alpha
+\beta-\sqrt{\left(  \alpha-\beta \right)  ^{2}+\mathrm{\gamma}_{\Vert}^{2}}
\label{mk}%
\end{equation}
Notice that using the condition $\mathrm{\gamma}_{\Vert}^{2}<4\mathrm{\alpha
\beta}$ and positivity of $\mathrm{\alpha}$ and $\mathrm{\beta}$, we have
$\left(  \alpha-\beta \right)  ^{2}+\mathrm{\gamma}_{\Vert}^{2}<\left(
\alpha+\beta \right)  ^{2}$ and then $\mathrm{\kappa}>0$\textrm{. }Notice also
the following properties:

\begin{itemize}
\item \textrm{The saturated value }$(\mathrm{\gamma}_{\Vert}^{2})_{\max
}=4\mathrm{\alpha \beta}$; then ($\det G_{ij})_{\max}=0$ and $\left(
\mathrm{\kappa}\right)  _{\min}\rightarrow2\left(  \alpha+\beta \right)
\omega_{\min}^{2}=0$\textrm{.}

\item Classically, the hamiltonian (\ref{h}) is positive and bounded from
below,
\begin{equation}
\boldsymbol{h}\geq \boldsymbol{h}_{0}\qquad,\qquad \boldsymbol{h}_{0}=0
\end{equation}
This vanishing lower value $\boldsymbol{h}_{0}=0$ is important in the study of
$\mathcal{N}=2$ gauged supergravity in the rigid limit; since $\left \langle
\mathcal{V}_{class}\right \rangle =\boldsymbol{h}_{0}=0$ corresponds to an
exact $\mathcal{N}=2$ rigid supersymmetric phase. This property requires
$\vec{m}=\vec{\nu}=\vec{0}$.

\item By restricting $\vec{m}$ and $\vec{\nu}$ to the particular choice
eq(\ref{sh}), the free eq(\ref{h}) reduces to the hamiltonian of a
1-dimensional harmonic oscillator,
\begin{equation}
\boldsymbol{h}^{\left(  1D\right)  }=\frac{1}{2\mathrm{\mu}}\left(
m_{y}^{\prime}\right)  ^{2}+\frac{\mathrm{\kappa}}{2}\left(  \nu_{x}^{\prime
}\right)  ^{2}%
\end{equation}

\end{itemize}

In what follows, we use this simple expression to study harmonic fluctuations
of the FI couplings around the supersymmetric vacuum $m_{y}^{\prime}=\nu
_{x}^{\prime}=0$.

\subsubsection{Quantum effect}

The free\ iso-particle studied above is classical. However, like spin
$s=\frac{1}{2}$\ fermions in the real 3d space, the iso-particle has also
intrinsic degrees of freedom namely an isospin $I=\frac{1}{2}$, as shown on
table (\ref{is}), and a unit electric/magnetic charge $\vartheta$ given by
(\ref{si}). \newline Assuming the classical correspondence (\ref{13}) to also
hold at the quantum level in the iso-space $\mathbb{\hat{R}}^{3}$, it follows
that the fluctuations of the FI couplings may also be governed by $\left \vert
\Delta \vec{m}^{\prime}\right \vert \times \left \vert \Delta \vec{\nu}^{\prime
}\right \vert \gtrsim \hbar$ in same manner as for the usual Heisenberg
uncertainty $\left \vert \Delta x\right \vert \times \left \vert \Delta
p_{x}\right \vert \gtrsim \hbar$ which is expressed in terms of the usual phase
space coordinates $\left(  \vec{r},\vec{p}\right)  $. If one accepts this
assumption; then we cannot have exact $m_{y}^{\prime}=\nu_{x}^{\prime}=0$
since $\left \vert \Delta \nu_{x}^{\prime}\right \vert \times \left \vert \Delta
m_{y}^{\prime}\right \vert \gtrsim \hbar$; and so one expects $\mathcal{N}=2$
supersymmetry in rigid limit to be broken by quantum effect since ground state
energy is now positive definite
\begin{equation}
\left \langle \boldsymbol{h}_{\text{quant}}^{\left(  1D\right)  }\right \rangle
>0
\end{equation}
In what follows, we restrict our study to exhibiting this quantum behaviour
and to checking the breaking $\mathcal{N}=2\rightarrow$ $\mathcal{N}=0$. We
will return to study this feature in subsection 4.2 when isospin-orbit
coupling is switched on. There, we will also give details on the condition to
have partial breaking $\mathcal{N}=2\rightarrow$ $\mathcal{N}=1$.\newline The
quantum effect, due to fluctuations of $\vec{m}$ and $\vec{\nu}$ around the
supersymmetric ground state $\boldsymbol{h}_{0}=\left \langle \mathcal{V}%
\right \rangle $,\ is induced by quantum isotropic oscillations with discrete
energy\textrm{\ }$\hat{\epsilon}_{\left(  n_{x},n_{y},n_{z}\right)
}^{\parallel}=\epsilon_{n_{x}}^{\parallel}+\epsilon_{n_{y}}^{\parallel
}+\epsilon_{n_{z}}^{\parallel}$\textrm{\ and fundamental }oscillation
frequency
\begin{equation}
\omega_{\Vert}=\sqrt{\frac{\mathrm{\kappa}}{\mathrm{\mu}}}%
\end{equation}
By using (\ref{mk}), we have,%
\begin{equation}
\omega_{\Vert}^{2}=4\alpha \beta-\mathrm{\gamma}_{\Vert}^{2} \label{4}%
\end{equation}
Observe that because of the minus sign, this $\omega_{\Vert}$ vanishes for
those parameters $\mathrm{\alpha}$, $\mathrm{\beta}$, $\mathrm{\gamma}_{\Vert
}$ satisfying the degenerate condition $\mathrm{\gamma}_{\Vert}^{2}%
=4\alpha \beta$ which has been ruled out by the constraint eq(\ref{det}). To
illustrate the quantum effect for $\omega_{\Vert}>0$, we consider the
particular choice (\ref{sh}) bringing (\ref{h}) to a 1-dimensional quantum
oscillator with hamiltonian operator as,%
\begin{equation}
\boldsymbol{H}_{\Vert}^{\left(  1D\right)  }=\frac{\hbar \omega_{\Vert}}%
{2}\left[  \left(  \frac{\mathbf{\hat{m}}_{y}}{\sqrt{\mathrm{\mu \omega}%
_{\Vert}}}\right)  ^{2}+\left(  \mathbf{\hat{\nu}}_{x}\sqrt{\mathrm{\mu \omega
}_{\Vert}}\right)  ^{2}\right]
\end{equation}
It has a diagonal form $\frac{\hbar \omega_{\Vert}}{2}\left(  Y^{2}%
+X^{2}\right)  $, which by setting $A=\frac{X+iY}{\sqrt{2}}$, reads as usual
like
\begin{equation}
\boldsymbol{H}_{\Vert}^{\left(  1D\right)  }=\hbar \omega_{\Vert}\left(
A^{\dagger}A+\frac{1}{2}\right)  \label{p}%
\end{equation}
with $AA^{\dagger}-A^{\dagger}A=I$. The energy spectrum $\hat{\epsilon
}_{\left(  n_{x},n_{y},n_{z}\right)  }^{\parallel}$ reduces to
\begin{equation}
\epsilon_{n}^{\parallel}=\hbar \omega_{\Vert}\left(  n+\frac{1}{2}\right)
\geq \epsilon_{0}%
\end{equation}
with frequency $\omega_{\Vert}$ given by (\ref{4}). The lowest energy value is
given by $\epsilon_{0}^{\parallel}=\frac{\hbar \omega_{\Vert}}{2}$; it is non
zero for non vanishing frequency\ $\omega_{\Vert}$. Hence, the exact
$\mathcal{N}=2$ supersymmetry living at classical vacuum $\left \langle
\mathcal{V}_{class}\right \rangle =0$ gets completely broken by quantum effect
\begin{equation}
\left \langle \mathcal{V}_{\text{quant}}\right \rangle =\frac{\hbar \omega
_{\Vert}}{2}>0
\end{equation}
We end this subsection by giving two brief comments on interactions. The first
interacting potential energy has the linear expression in $\vec{m}$,
\begin{equation}
h_{int}^{\left(  \hat{R}^{3}\right)  }=-q\vec{m}.\mathcal{\vec{A}} \label{S}%
\end{equation}
and concerns the electric $U\left(  1\right)  _{elec}$ gauge charge. This is a
subgroup of the electric/magnetic $U\left(  1\right)  _{elec}\times U\left(
1\right)  _{mag}$ local symmetry of the $\mathcal{N}=2$\ gauged supergravity
induced by gauging two abelian isometries in the scalar manifold of the
supergravity theory. The second interacting potential energy is given by the
isospin-orbit coupling $h_{ioc}^{\left(  \hat{R}^{3}\right)  }=\vec{\xi
}.\mathcal{\vec{I}}$ we are too particularly interested in here; it will be
considered with details in the next subsection. \newline Regarding (\ref{S}),
it is derived by making two steps as follows: First, start from the
interaction energy $h_{int}^{\left(  R^{3}\right)  }=-e\vec{p}.\vec{A}$ of an
electrically charged particle with momentum $\vec{p}$ moving in the presence
of an external magnetic field $\vec{B}_{ext}=\vec{\nabla}\wedge \vec{A}$. Then,
use the correspondence (\ref{13}) allowing to imagine $-e\vec{p}$ in terms of
the FI magnetic vector $-q\vec{m}$ and the $\vec{A}$ by an iso-vector
$\mathcal{\vec{A}}$. The obtained (\ref{S}) describes just the term
$w_{a}m^{a}$ in eq(\ref{xe}) from which we learn that $\vec{w}=-q\mathcal{\vec
{A}}$.

\subsection{Isospin- orbit coupling}

The proposal (\ref{13}) has been useful for the physical interpretation of the
rigid Ward identity in terms of an iso-particle hamiltonian with phase space
coordinates $\left(  \vec{\nu},\vec{m}\right)  $. Thanks to the Andrianoploli
et \underline{al} formula $\vec{\xi}=\vec{\nu}\wedge \vec{m}$ giving the
orbital momentum of this iso-particle. Thanks also to the structure of the
scalar potential $\mathcal{V}$ which turns out to be nothing but its the free
hamiltonian $\boldsymbol{h}$ (\ref{h}). In this subsection, we derive the
isospin-orbit coupling%
\begin{equation}
\boldsymbol{h}_{ioc}=\vec{\xi}.\mathcal{\vec{I}} \label{io}%
\end{equation}
where $\mathcal{\vec{I}}$ stands for the isospin vector and $\vec{\xi}%
=\vec{\nu}\wedge \vec{m}$. To that purpose, recall that in eq(\ref{9b}), the
rigid $\boldsymbol{C}$\emph{- }anomaly matrix appears in the form of a
hermitian traceless 2$\times$2 matrix
\begin{equation}
\boldsymbol{C}=\left(
\begin{array}
[c]{cc}%
\xi_{z} & \xi_{x}-i\xi_{y}\\
\xi_{x}+i\xi_{y} & -\xi_{z}%
\end{array}
\right)  =\vec{\xi}.\vec{\tau}%
\end{equation}
that reads in terms of the $\vec{\tau}$- Pauli matrices and the Andrianopoli
et \underline{al} orbital vector as follows%
\begin{equation}
\boldsymbol{C}=\left(  \vec{\nu}\wedge \vec{m}\right)  .\vec{\tau}%
\end{equation}
This factorised form of $\boldsymbol{C}$ teaches us that it can be imagined as
describing the coupling of two things namely the orbital isovector $\vec{\xi
}=\vec{\nu}\wedge \vec{m}$ and the isospin vector
\begin{equation}
\mathcal{\vec{I}}=\frac{\vec{\tau}}{2}%
\end{equation}
In what follows, we give two other different, but equivalent, manners to
introduce $\vec{\xi}.\mathcal{\vec{I}}$. The first way relies on comparing
$\boldsymbol{h}_{ioc}=\vec{\xi}.\mathcal{\vec{I}}$ with the usual spin-orbit
coupling $\boldsymbol{h}_{soc}=\vec{L}.\vec{S}$ of a particle with spin
$\vec{S}=\frac{\vec{\sigma}}{2}$ moving in real space $\mathbb{R}^{3}$ with
coordinate vector $\vec{r}$. The second manner extends the approach done in
previous section for deriving the free hamiltonian (\ref{h}) by including the
isospin effect. \newline By comparing the effect of the spin orbit coupling
$\vec{L}.\vec{S}$ in electronic systems and the effect of $\vec{\xi
}.\mathcal{\vec{I}}$ in partial breaking of $\mathcal{N}=2$ supersymmetry; and
by following \textrm{\cite{12,13}}, we learn that when the central extension
matrix is turned off, i.e: $\boldsymbol{C}=0$, then $\mathcal{N}=2$
supersymmetry is preserved (two gapless gravitinos). However, it can be
partially broken when it is turned on i.e: $\boldsymbol{C}\neq0$. This
property can be viewed in terms of a non zero gap energy E$_{g}$ between the
two fermionic iso-doublets; including the two charges $Q_{L},Q_{R}$ of
$\mathcal{N}=2$ supersymmetry with expression as%
\begin{equation}
E_{g}\propto \left \vert \vec{\xi}\right \vert
\end{equation}
This is exactly what happens for the case of two states of spin $\frac{1}{2}$
fermions in electronic condensed matter systems when the spin- orbit coupling
$\vec{L}.\vec{S}$ is taken into account. This $\vec{L}.\vec{S}$ coupling is
known to open the zero gap between the two states of free electrons. From this
link with electronic properties, we deduce a correspondence between the
central matrix $\boldsymbol{C}$ of the $\mathcal{N}=2$ supercurrent algebra
and the hamiltonian $\boldsymbol{h}_{soc}=\vec{L}.\vec{S}$. This link reads
explicitly like
\begin{equation}
\vec{\xi}.\mathcal{\vec{I}}\qquad \leftrightarrow \qquad \vec{L}.\vec{S}%
\end{equation}
where the isospin $\mathcal{\vec{I}}$ plays the role of the spin $\vec{S}$;
and the Andrianopoli et \underline{al} vector $\vec{\xi}$ the role of the
angular momentum $\vec{L}$. Adding the isospin- orbit coupling term to the
free (\ref{h}), we get $H=\mathcal{V}+\vec{\xi}.\mathcal{\vec{I}}$ that reads
explicitly as follows%
\begin{equation}
H=\frac{1}{2\mathrm{\mu}}\vec{m}^{\prime2}+\frac{\mathrm{\kappa}}{2}\vec{\nu
}^{\prime2}+\vec{\xi}.\mathcal{\vec{I}} \label{hh}%
\end{equation}
In matrix form, we have%
\begin{equation}
H=\left(
\begin{array}
[c]{cc}%
\mathcal{V}+\xi_{z} & \xi_{x}-i\xi_{y}\\
\xi_{x}+i\xi_{y} & \mathcal{V}-\xi_{z}%
\end{array}
\right)  \label{hg}%
\end{equation}
with eigenvalues: $E_{\pm}=\mathcal{V}\pm \sqrt{\xi_{x}^{2}+\xi_{y}^{2}+\xi
_{z}^{2}}$ and eigenstates as%
\begin{equation}
\left \vert \eta_{\pm}\right \rangle \sim \left(
\begin{array}
[c]{c}%
\xi_{z}\pm \sqrt{\xi_{x}^{2}+\xi_{y}^{2}+\xi_{z}^{2}}\\
\xi_{x}+i\xi_{y}%
\end{array}
\right)  \label{eta}%
\end{equation}
The second manner to introduce (\ref{io}) is a purely algebraic approach. The
key idea relies on thinking of the free energy density term of the two
$I_{z}=\pm \frac{1}{2}$\ isospin states as
\begin{equation}
\boldsymbol{h}_{B}^{A}=\mathcal{V}\delta_{B}^{A}%
\end{equation}
To each of the $I_{z}=\pm \frac{1}{2}$ states, we have used (\ref{h}) to derive
its free hamiltonian; but this result is just the diagonal term of a general
hamiltonian matrix $H$. The extension of $\mathcal{V}\delta_{B}^{A}$ to have
more interactions is then naturally given by the Ward identity (\ref{9b})
\begin{equation}
\boldsymbol{H}_{B}^{A}=\mathcal{V}\delta_{B}^{A}+C_{B}^{A}%
\end{equation}
which is nothing but the right hand side of the $\mathcal{N}=2$ supersymmetric
current algebra (\ref{ac}) including the central matrix.

\subsection{Discrete symmetries}

From the rigid Ward identity of Andrianopoli et \underline{al} (\ref{9b}), we
also learn that exact $\mathcal{N}=2$ supersymmetry requires $\vec{\xi}%
=\vec{0}$; no isospin-orbit coupling in our modeling. But this vanishing value
is just the fix point of the $\boldsymbol{Z}_{2}$ discrete symmetry acting on
the anomaly isovector as follows
\begin{equation}
\boldsymbol{Z}_{2}:\quad \vec{\xi}\quad \rightarrow \quad-\vec{\xi} \label{ksi}%
\end{equation}
To figure out the meaning of this discrete transformation, we use
eq(\ref{10b}) from which we learn that the minus sign can be generated in two
manners; either by the change $\left(  \vec{\nu},\vec{m}\right)
\rightarrow \left(  -\vec{\nu},\vec{m}\right)  $; or by $\left(  \vec{\nu}%
,\vec{m}\right)  \rightarrow \left(  \vec{\nu},-\vec{m}\right)  $. To derive
the physical interpretation to these two kinds of $\boldsymbol{Z}_{2}$
discrete symmetries, we use the analogy between the FI couplings $\left(
\vec{\nu},\vec{m}\right)  $ and the classical phase coordinates $\left(
\vec{r},\vec{p}\right)  $. Promoting this correspondence to dynamical
(running) couplings; say
\begin{equation}
\left.
\begin{array}
[c]{c}%
\vec{r}\left(  t\right) \\
\vec{p}\left(  t\right)
\end{array}
\right.  \qquad \leftrightarrow \qquad \left.
\begin{array}
[c]{c}%
\vec{\nu}\left(  \mathrm{\tau}\right) \\
\vec{m}\left(  \mathrm{\tau}\right)
\end{array}
\right.  \label{cr}%
\end{equation}
it follows that the transformation (\ref{ksi}) corresponds for example to the
usual time reversing symmetry $\boldsymbol{T}$ which maps the position
$\vec{r}\left(  t\right)  $ and momentum $\vec{p}\left(  t\right)  $
respectively to $\vec{r}\left(  -t\right)  $ and $-\vec{p}\left(  -t\right)
$. On the side of the FI couplings, we then have the following action of the
$\boldsymbol{T} $- analogue on isotime $\mathrm{\tau}$,%
\begin{equation}
\boldsymbol{T}:\left.
\begin{array}
[c]{c}%
\vec{\nu}\left(  \mathrm{\tau}\right) \\
\vec{m}\left(  \mathrm{\tau}\right)
\end{array}
\right.  \qquad \rightarrow \qquad \left.
\begin{array}
[c]{c}%
\vec{\nu}\left(  -\mathrm{\tau}\right) \\
-\vec{m}\left(  -\mathrm{\tau}\right)
\end{array}
\right.
\end{equation}
Notice that the usual space parity $\boldsymbol{P}$ which maps the $\left(
\vec{r},\vec{p}\right)  $ phase coordinates to $\left(  -\vec{r},-\vec
{p}\right)  $ allows us, by using the $\left(  \vec{r},\vec{p}\right)
\leftrightarrow \left(  \vec{\nu},\vec{m}\right)  $ correspondence, to write
\begin{equation}
\boldsymbol{P}:\left.
\begin{array}
[c]{c}%
\vec{\nu}\left(  \mathrm{\tau}\right) \\
\vec{m}\left(  \mathrm{\tau}\right)
\end{array}
\right.  \qquad \rightarrow \qquad \left.
\begin{array}
[c]{c}%
-\vec{\nu}\left(  \mathrm{\tau}\right) \\
-\vec{m}\left(  \mathrm{\tau}\right)
\end{array}
\right.
\end{equation}
But this discrete $\boldsymbol{P}$- transformation leaves $\vec{\xi}=\vec{\nu
}\wedge \vec{m}$ invariant and so it is not relevant for partial breaking.
However, the combined $\boldsymbol{PT}$ transformation which acts like%
\begin{equation}
\boldsymbol{PT}:\left.
\begin{array}
[c]{c}%
\vec{\nu}\left(  \mathrm{\tau}\right) \\
\vec{m}\left(  \mathrm{\tau}\right)
\end{array}
\right.  \qquad \rightarrow \qquad \left.
\begin{array}
[c]{c}%
-\vec{\nu}\left(  -\mathrm{\tau}\right) \\
+\vec{m}\left(  -\mathrm{\tau}\right)
\end{array}
\right.  \label{ks}%
\end{equation}
does affect the sign of $\vec{\xi}$. This combination can be also used to
think about the $\boldsymbol{Z}_{2}$ transformation (\ref{ksi}). Actually, it
corresponds to the second possibility to realize $\vec{\xi}\rightarrow
-\vec{\xi}$ from (\ref{10b}). Therefore, exact $\mathcal{N}=2$ supersymmetry,
which corresponds to $\vec{\xi}=\vec{0}$, lives at the fix point of the
$\boldsymbol{T}$- reversing time transformation (\ref{ksi}); or at the
combined $\boldsymbol{PT}$ given by (\ref{ks}); or both.

\section{Topological aspects and quantum effect}

In this section, we first study the topological behaviour of gapless
iso-particles of exact $\mathcal{N}=2$ supersymmetry as well as the gapless
chiral ones that remain after partial breaking. Then, we study the effect of
quantum fluctuations on partial supersymmetry breaking.

\subsection{Chiral anomaly}

\begin{figure}[ptbh]
\begin{center}
\hspace{2cm}\includegraphics[scale=0.6]{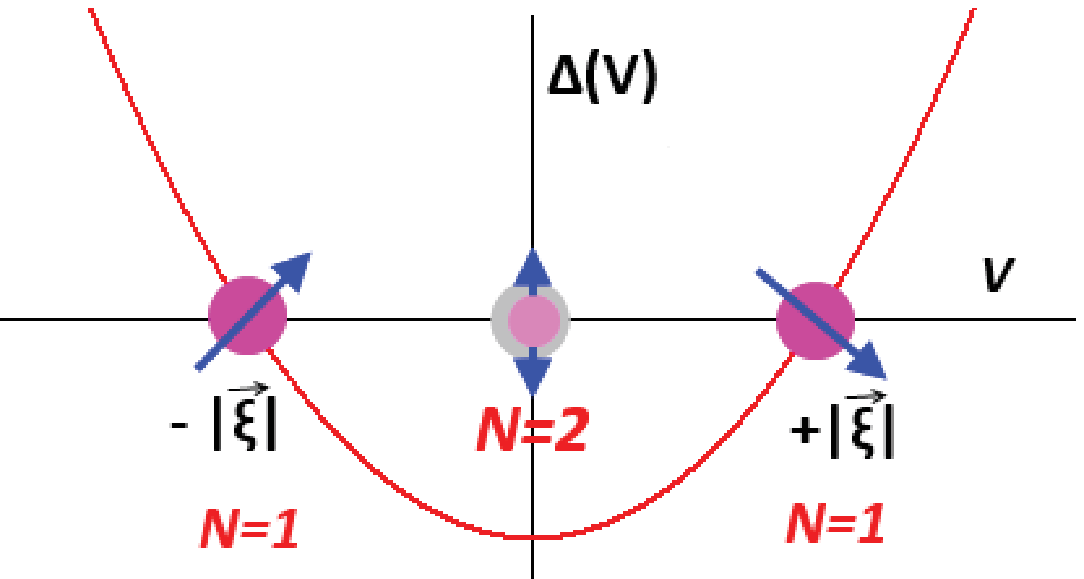}\quad
\includegraphics[scale=0.6]{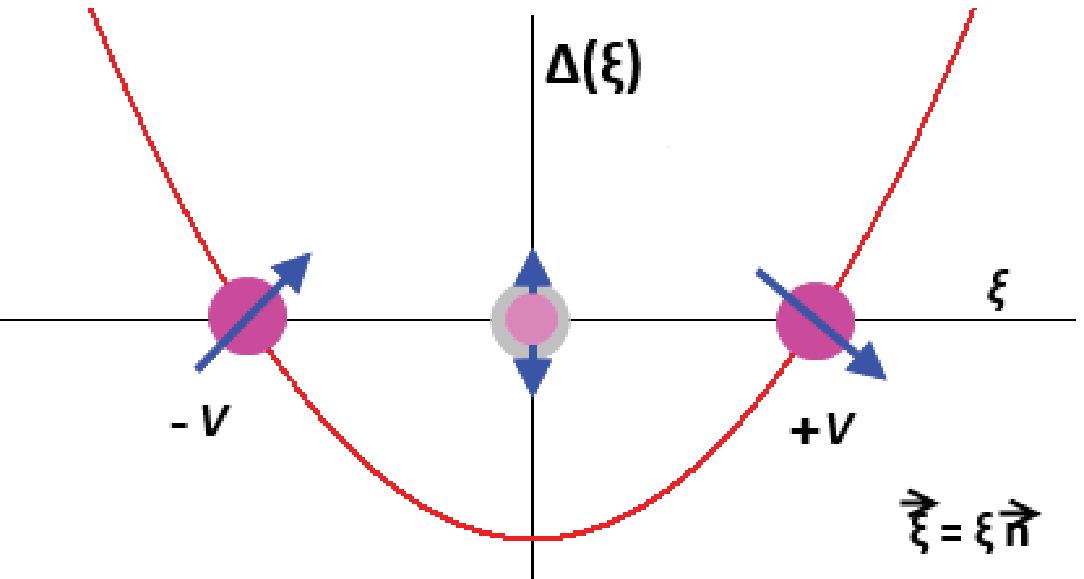}
\end{center}
\par
\vspace{1cm} \caption{On left, the discriminant $\Delta_{\xi}\left(  V\right)
$ as a function of $V$ and parameter $\left \vert \vec{\xi}\right \vert $. For
$\vec{\xi}\neq \vec{0};$ there are two zeros at $V=\pm \left \Vert \vec{\xi
}\right \Vert $; one visible in rigid limit. At each zero, say $V=\left \Vert
\vec{\xi}\right \Vert $; it lives a chiral gapless corresponding to a partially
broken $\mathcal{N}=2$ supersymmetric state. In the limit $\vec{\xi
}\rightarrow \vec{0}$, the two chiral gapless modes at $V=\pm \left \Vert
\vec{\xi}\right \Vert $ collide at the origin and form together a gapless
isodoublet. On right, the $\Delta_{V}\left(  \xi \right)  $ as a function of
$\vec{\xi}$. For non zero $V;$ chiral gapless mode live at each $\vec{\xi
}_{\pm}=\pm V\  \vec{n}$ merging for $V=0$. \qquad \qquad \qquad}%
\label{2}%
\end{figure}

Setting $H_{B}^{A}=\sum_{i}\delta_{B}\lambda^{iC}\delta^{A}\lambda_{iC}$, we
can turn the rigid Ward identities (\ref{9b}) into the matrix equation
\begin{equation}
H_{B}^{A}=\mathcal{V}\delta_{B}^{A}+C_{B}^{A}\label{hab}%
\end{equation}
which is nothing but the hamiltonian matrix (\ref{hh}). Multiplying both sides
of this 2$\times$2 matrix relation by $\eta_{A}=\left(  \eta_{1},\eta
_{2}\right)  ^{T}$ describing the two states of the isoparticle, we end with
the eigenvalue equation $H.\eta=E\eta$ whose two eigenvalues are given by
$E_{\pm}=\mathcal{V}\pm \left \vert \vec{\xi}\right \vert $; the eigenstates
$\hat{\eta}_{\pm}$ associated with these $E_{\pm}$ are linear combinations of
$\eta_{1}$ and $\eta_{2}$, they read like\textrm{ }$\hat{\eta}_{\pm}=A_{\pm
}\eta_{1}+B_{\pm}\eta_{2}$ with amplitudes $A_{\pm}$ and $B_{\pm}$ as follows%
\begin{equation}
A_{\pm}=\frac{\xi_{z}\pm \left \vert \vec{\xi}\right \vert }{\sqrt{2\left(
\xi_{z}-\left \vert \vec{\xi}\right \vert \right)  }}\qquad,\qquad B_{\pm}%
=\frac{\xi_{x}+i\xi_{y}}{\sqrt{2\left(  \xi_{z}-\left \vert \vec{\xi
}\right \vert \right)  }}%
\end{equation}
\textrm{ The determ}inant $\det H=\Delta$ that captures data on the singular
points in the ($\mathcal{V},\left \vert \vec{\xi}\right \vert $) plane is given
by the product of the eigenvalues $E_{\pm}$, it reads as
\begin{equation}
\Delta=\left(  \mathcal{V}+\left \vert \vec{\xi}\right \vert \right)  \left(
\mathcal{V}-\left \vert \vec{\xi}\right \vert \right)  \label{ml}%
\end{equation}
It is a function of two real quantities namely $\mathcal{V}$ and $\left \vert
\vec{\xi}\right \vert $; but here below we will treated it as a parametric
function of one variable like $\Delta_{\zeta}\left(  x\right)  $. The choice
of the variable $x$ depends on the property we are interested to exhibit; see
figure (\ref{2}). From the view of the scalar potential energy, the variable
is given by $x=\mathcal{V}$ ; while $\zeta=\left \vert \vec{\xi}\right \vert $
is seen as a free parameter.\newline From the view of the $\vec{\xi}$ vector, we have the reverse
picture; $x=\left \vert \vec{\xi}\right \vert $ is the variable while
$\zeta=\mathcal{V}$ stands for a free parameter. In the first image, $\det H$
has two zeros at $\mathcal{V}_{\pm}=\pm \left \vert \vec{\xi}\right \vert $; one
positive $\mathcal{V}_{+}$, that is visible in global supersymmetry sector;
and a hidden negative $\mathcal{V}_{-}$. In the second picture, the
discriminant $\det H$ has zeros at $\left \vert \vec{\xi}_{+}\right \vert
=+\mathcal{V}$ for positive $\mathcal{V}$; and $\left \vert \vec{\xi}%
_{-}\right \vert =-\mathcal{V}$ for negative $\mathcal{V}$. Let us express
these two zeros in $\mathbb{\tilde{R}}^{3}$ like $\vec{\xi}_{\pm}%
=\pm \mathcal{V}\vec{n}$ with unit vector $\vec{n}=\frac{\vec{\xi}}{\left \vert
\vec{\xi}\right \vert }$. The effective gap energy $E_{g}=E_{+}-E_{-}$ between
the two $E_{\pm}$ energy density bands is given by
\begin{equation}
E_{g}=2\left \vert \vec{\xi}\right \vert
\end{equation}
It vanishes for $\left \vert \vec{\xi}\right \vert =0$ and then for $\vec{\xi
}=\vec{0}$. Because of the property $\left \vert \vec{\xi}\right \vert \geq0$,
the zeros of $\det H$ are of two kinds: simple for $\left \vert \vec{\xi
}\right \vert >0$ and double for $\left \vert \vec{\xi}\right \vert =0$. At each
simple zero lives a gapless fermionic mode (gravitino and gaugino) and a
gapped one. For $\left \vert \vec{\xi}_{+}\right \vert =+\mathcal{V}$ with
positive energy density $\mathcal{V}$, we have the conducting band; and for
$\left \vert \vec{\xi}_{-}\right \vert =-\mathcal{V}$ with negative
$\mathcal{V}$, we have the valence band. Notice that $\det H=\mathcal{V}%
^{2}-\vec{\xi}^{2}$ is conserved under\textrm{\footnote{\ from charged
particles view, this mapping from valence to conducting like bands and vice
versa may be imagined as a $\boldsymbol{CT}$ transformation combining time
reversing $\boldsymbol{T}$ and charge conjugation $\boldsymbol{C}$. } }the
discrete change\textrm{,}
\begin{equation}
\boldsymbol{Z}_{2}:\mathcal{V}\rightarrow-\mathcal{V}\qquad \Rightarrow
\qquad \vec{\xi}_{+}\rightarrow \vec{\xi}_{-}=-\vec{\xi}_{+}\label{fx}%
\end{equation}
Its two zeros $\mathcal{V}_{\pm}=\pm \left \vert \vec{\xi}\right \vert $ are not
fix points of $\boldsymbol{Z}_{2}$ except the origin; they are interchanged as
shown by (\ref{fx}); for instance properties at $\vec{\xi}_{-}$ may be deduced
from those at $\vec{\xi}_{+}$. \newline Now, let us approach $\det H$ from the
view of the iso-space vector $\vec{\xi}$; and consider the 2-spheres
$\mathbb{S}_{+\upsilon}^{2}$ and $\mathbb{S}_{-\upsilon}^{2}$, with a surface
normal to $\vec{n}$, surrounding respectively the zeros
\begin{equation}
\vec{\xi}_{\pm}=\pm \mathcal{V}\vec{n}\label{ck}%
\end{equation}
The 2-sphere $\mathbb{S}_{+\upsilon}^{2}$ is described by the vector $\vec
{p}=\vec{\xi}-\vec{\xi}_{+}$ and the $\mathbb{S}_{-\upsilon}^{2}$ by $\vec
{q}=\vec{\xi}-\vec{\xi}_{-}$. These 2-spheres should not be confused with the
unit 2-sphere
\begin{equation}
\mathbb{S}_{\vec{n}}^{2}:n_{x}^{2}+n_{y}^{2}+n_{z}^{2}=1
\end{equation}
associated with the unit vectors (\ref{20}); but the three $\mathbb{S}%
_{+\upsilon}^{2}$, $\mathbb{S}_{-\upsilon}^{2}$, $\mathbb{S}_{\vec{n}}^{2}$
live all of them in the iso-space $\mathbb{\tilde{R}}^{3}$; and are related to
each other by continuous mappings like,%
\begin{equation}
\pi_{+}:\mathbb{S}_{+\upsilon}^{2}\rightarrow \mathbb{S}_{\vec{n}}^{2}%
\qquad,\qquad \pi_{-}:\mathbb{S}_{-\upsilon}^{2}\rightarrow \mathbb{S}_{\vec{n}%
}^{2}\label{il}%
\end{equation}
Focussing for instance on $\mathbb{S}_{+\upsilon}^{2}$, the continuity of
$\pi_{+}$ shows that it has a winding $w(\mathbb{S}_{+\upsilon}^{2})$
describing the net number of times $\mathbb{S}_{+\upsilon}^{2}$ wraps the unit
sphere $\mathbb{S}_{\vec{n}}^{2}$; the integer number $w(\mathbb{S}%
_{+\upsilon}^{2})$ reflects just the mathematical property $\pi_{2}\left(
\mathbb{S}^{2}\right)  \cong \mathbb{Z}$. A similar thing can be said about
$\mathbb{S}_{-\upsilon}^{2}$; thanks to $\boldsymbol{Z}_{2}$ parity (\ref{fx})
under which gauge curvature $\mathcal{F}$ of underlying Berry connection
$\mathcal{A}$ is odd; see (\ref{fa}) given below.\newline Moreover, each one
of gapless state at the two zeros $\vec{\xi}_{\pm}=\pm \mathcal{V}\vec{n}$ is
anomalous in the sense that it has one gapless chiral mode and then violates
the Nielson-Ninomiya theorem \textrm{\cite{21,210,6}}. Recall that in theories
that are free from chiral anomalies, the usual Nielson- Ninomiya theorem
\textrm{\cite{21,6}} states that the sum of winding numbers $w\left(
\mathbb{S}_{i}^{2}\right)  $ around 2-spheres $\mathbb{S}_{i}^{2}$,
surrounding the $\vec{\xi}_{\ast i}$ zeros where live gapless modes, vanishes
identically. Here, this statement reads explicitly like%
\begin{equation}
\sum_{i}w\left(  S_{i}^{2}\right)  =\sum_{i}\int_{\mathbb{S}_{i}^{2}}%
\frac{Tr\left(  \mathcal{F}\right)  }{2\pi}=0
\end{equation}
where $\mathcal{F}$ is a gauge curvature whose explicit expression will be
given below. For positive $\mathcal{V}$, eq(\ref{ml}) has one zero given by an
outgoing $\vec{\xi}_{+}=+\mathcal{V}\vec{n}$ with positive sense in normal
$\vec{n}$ direction; and then a 2-sphere $\mathbb{S}_{+\upsilon}^{2}$
surrounding the point $\vec{\xi}_{+}=\left(  \xi_{+x},\xi_{+y},\xi
_{+z}\right)  $ has a positive winding number
\begin{equation}
w\left(  S_{+}^{2}\right)  =\int_{\mathbb{S}_{+\upsilon}^{2}}\frac{Tr\left(
\mathcal{F}\right)  }{2\pi}=1
\end{equation}
Here, the curvature $\mathcal{F}$ is given by the following rank 2
antisymmetric tensor,
\begin{equation}
\mathcal{F}_{ab}=\frac{1}{2}\vec{n}.\left(  \frac{\partial \vec{n}}{\partial
\xi^{a}}\wedge \frac{\partial \vec{n}}{\partial \xi^{b}}\right)  \label{fa}%
\end{equation}
The Nielson- Ninomiya theorem is then violated due to the existence of one
gapless chiral moving mode; and so the partially broken theory has a chiral
anomaly; only one of the two supersymmetric charges $\left(  \hat{Q}_{L}%
,\hat{Q}_{R}\right)  $; say the right $\hat{Q}_{R}$ is preserved; the left
$\tilde{Q}_{L}$ is broken. For the incoming $\vec{\xi}_{-}=-\mathcal{V}\vec
{n}$, we have a negative winding number%
\begin{equation}
w\left(  S_{-}^{2}\right)  =\int_{\mathbb{S}_{-\upsilon}^{2}}\frac{Tr\left(
\mathcal{F}\right)  }{2\pi}=-1
\end{equation}
This negative value follows from the mapping $\vec{n}\rightarrow-\vec{n}$, due
to eq(\ref{fx}), and using (\ref{fa}). For the special case where the VEV of
the scalar potential vanish, $\mathcal{V}=0,$ the discriminant of the matrix
(\ref{ml}) reduces to $\det \boldsymbol{H}=-\left \vert \vec{\xi}\right \vert
^{2}$ and its zero $\left \vert \vec{\xi}\right \vert =0$ has a multiplicity 2.
In this case, the Nielson-Ninomiya theorem reads as
\begin{equation}
w\left(  S_{+}^{2}\right)  +w\left(  S_{-}^{2}\right)  =1-1=0
\end{equation}
At the fix point of the transformation (\ref{fx}), the two zeros collide at
$\left \vert \vec{\xi}_{\pm}\right \vert =0$. Then, the two effective gravitino
zero modes with opposite chiralities form a massless doublet ( a massless
iso-particle) and $\mathcal{N}=2$ supersymmetry gets restored.%
\begin{equation}%
\begin{tabular}
[c]{l|l|l|l}%
{\small zeros of} $\det H$ \  \  \  & \ {\small multiplicity of zeros \ } &
\ {\small winding number \ } & \ {\small conserved SUSY charges \ }\\ \hline
$\  \  \  \left \vert \vec{\xi}_{+}\right \vert {\small =+}\mathcal{V}$ &
${\small \  \  \  \  \  \  \  \  \  \  \  \  \  \ 1}$ & ${\small \  \  \  \  \  \ +1}$ &
${\small \  \  \  \  \  \  \  \  \  \  \  \  \tilde{Q}}^{+}\left.
\begin{array}
[c]{c}%
\text{ \ }\\
\text{ \ }%
\end{array}
\right.  $\\
$\  \  \  \left \vert \vec{\xi}_{-}\right \vert {\small =-}\mathcal{V}$ &
${\small \  \  \  \  \  \  \  \  \  \  \  \  \  \ 1}$ & ${\small \  \  \  \  \  \ -1}$ &
${\small \  \  \  \  \  \  \  \  \  \  \  \  \tilde{Q}}^{-}$\\
$\  \  \  \left \vert \vec{\xi}_{\pm}\right \vert {\small =0}$ &
${\small \  \  \  \  \  \  \  \  \  \  \  \  \  \ 2}$ & ${\small \  \  \  \  \  \  \  \ 0}$ &
${\small \  \  \  \  \  \  \ }\left(
\begin{array}
[c]{c}%
{\small \tilde{Q}}^{+}\\
{\small \tilde{Q}}^{-}%
\end{array}
\right)  $\\ \hline
\end{tabular}
\  \  \  \  \label{t}%
\end{equation}

\subsection{Quantum fluctuation}

Here, we study quantum fluctuations in the FI couplings around the partial
breaking vacuum $\left \langle \mathcal{V}\right \rangle =\left \vert \vec{\xi
}\right \vert $ and comment on their effect by using the special choice
(\ref{sh}). To that purpose, we use $\left \vert \Delta \vec{m}^{\prime
}\right \vert \times \left \vert \Delta \vec{\nu}^{\prime}\right \vert \sim \hbar$
to promote the matrix equation (\ref{hab}) into an effective quantum
eigenvalue matrix equation $\boldsymbol{H}\left \vert \eta \right \rangle
=E\left \vert \eta \right \rangle $ that we split into two eigenvalues equations
as follows%
\begin{equation}%
\begin{tabular}
[c]{lll}%
$\boldsymbol{H}_{+}\left \vert \eta_{+}\right \rangle $ & $=$ & $E_{+}\left \vert
\eta_{+}\right \rangle $\\
$\boldsymbol{H}_{-}\left \vert \eta_{-}\right \rangle $ & $=$ & $E_{-}\left \vert
\eta_{-}\right \rangle $%
\end{tabular}
\  \  \label{pm}%
\end{equation}
In these relations, we have $\boldsymbol{H}_{\pm}=\mathcal{\hat{V}}%
\pm \mathbf{\hat{\xi}}$ where the hatted $\mathcal{\hat{V}}$ and $\mathbf{\hat
{\xi}}$ refer to the quantised operators associated with $\mathcal{V}$ and
$\left \vert \vec{\xi}\right \vert $ expressed in terms of the phase space
vectors $\vec{m}$ and $\vec{\nu}$. For the particular FI coupling choice
(\ref{sh}), we have $\left \vert \Delta m_{y}^{\prime}\right \vert
\times \left \vert \Delta \nu_{x}^{\prime}\right \vert \sim \hbar$ and find, after
repeating the steps between eqs(\ref{hp}) and (\ref{p}), the two following
quantum 1D- hamiltonians%
\begin{equation}
\boldsymbol{H}_{\pm}^{\left(  1D\right)  }=\hbar \omega_{\pm}\left(
A^{\dagger}A+\frac{1}{2}\right)
\end{equation}
describing two oscillators with different frequencies $\omega_{\pm}$. Their
energies are given by $\epsilon_{n}^{\pm}=\hbar \omega_{\pm}\left(  n+\frac
{1}{2}\right)  $ with%
\begin{equation}
\omega_{\pm}^{2}=4\alpha \beta-\left(  \mathrm{\gamma}_{\Vert}\pm
\mathrm{\gamma}_{\perp}\right)  ^{2} \label{mo}%
\end{equation}
with the remarkable minus sign. Notice that imposing the constraint
(\ref{det}) to both $\left \vert \eta_{\pm}\right \rangle $ eigenstates, we
have
\begin{equation}
\mathrm{\alpha \beta-}\frac{\left(  \mathrm{\gamma}_{\Vert}-\mathrm{\gamma
}_{\perp}\right)  ^{2}}{4}\geq0\mathrm{\qquad},\qquad \mathrm{\alpha \beta
-}\frac{\left(  \mathrm{\gamma}_{\Vert}+\mathrm{\gamma}_{\perp}\right)  ^{2}%
}{4}\geq0
\end{equation}
leading to%
\begin{equation}
0\leq \left(  \mathrm{\gamma}_{\Vert}-\mathrm{\gamma}_{\perp}\right)  ^{2}%
\leq4\mathrm{\alpha \beta \qquad},\qquad0\leq \left(  \mathrm{\gamma}_{\Vert
}+\mathrm{\gamma}_{\perp}\right)  ^{2}\leq4\mathrm{\alpha \beta} \label{tb}%
\end{equation}
and then to
\begin{equation}
-\mathrm{\alpha \beta}\leq \mathrm{\gamma}_{\Vert}\mathrm{\gamma}_{\perp}%
\leq \mathrm{\alpha \beta}%
\end{equation}
For the case where one of the bounds of the constraint eqs(\ref{tb}) is
saturated; for example the upper bound of the squared deviation
$(\mathrm{\gamma}_{\Vert}-\mathrm{\gamma}_{\perp})^{2}\leq4\mathrm{\alpha
\beta}$ is saturated, we can fix one of the four parameters in terms of the
three others like
\begin{equation}
\left(  \mathrm{\gamma}_{\Vert}-\mathrm{\gamma}_{\perp}\right)  ^{2}%
=4\mathrm{\alpha \beta \qquad}\Rightarrow \qquad \mathrm{\gamma}_{\perp
}=\mathrm{\gamma}_{\Vert}\pm2\sqrt{\mathrm{\alpha \beta}} \label{sc}%
\end{equation}
By substituting back into (\ref{mo}), we end with the two energy spectrums:
First, $\epsilon_{n}^{-}=\hbar \omega_{-}^{sat}\left(  n+\frac{1}{2}\right)  $
with%
\begin{equation}
\omega_{-}^{sat}=4\alpha \beta-\left(  \mathrm{\gamma}_{\Vert}-\mathrm{\gamma
}_{\perp}\right)  ^{2}=0 \label{sat}%
\end{equation}
describing gapless iso-particles (gravitinos/gauginos) with $E_{-}=0$. This
corresponds to the ground state $\left \langle \mathcal{V}\right \rangle
=\left \vert \vec{\xi}\right \vert $ where partial breaking takes place. Second,
$\epsilon_{n}^{+}=\hbar \omega_{+}^{sat}\left(  n+\frac{1}{2}\right)  $ with%
\begin{equation}
\left(  \omega_{+}^{sat}\right)  ^{2}=4\alpha \beta-\left(  \mathrm{\gamma
}_{\Vert}+\mathrm{\gamma}_{\perp}\right)  ^{2}=-4\mathrm{\gamma}_{\Vert
}\mathrm{\gamma}_{\perp}>0 \label{m}%
\end{equation}
They describe a gapped iso-particle. Thus, along with the gapless modes
($\omega_{-}^{sat}=0)$, we have gapped states with harmonics $n\omega
_{+}^{sat}$. The $\epsilon_{n}^{+}$ energies are bounded as
\begin{equation}
\epsilon_{n}^{+}\geq \epsilon_{0}^{+}=\frac{1}{2}\hbar \omega_{+}^{sat}>0
\end{equation}
with ground state energy $\epsilon_{0}^{+}$ corresponding to the classical
$E_{+}=2\left \vert \vec{\xi}\right \vert $; which is also the gap energy
between the two polarisations of the iso-particle. As a conclusion of this
subsection, quantum fluctuations in the FI coupling space with $\mathrm{\gamma
}_{\perp}=\mathrm{\gamma}_{\Vert}\pm2\sqrt{\mathrm{\alpha \beta}}$ do not
destroy the partial breaking supersymmetry of Andrianopoli et \underline{al}
rigid limit; this property holds for the saturated condition (\ref{sc});
otherwise quantum corrections break as well the residual $\mathcal{N}=1$ supersymmetry.

\section{Conclusion}

In this paper, we have used results on topological band theory of usual
$s=\frac{1}{2}$ matter to study partial breaking of $\mathcal{N}=2$ gauged
supergravity in rigid limit. By using supergravity Ward identities and results
from \textrm{\cite{12} and \cite{6,7,8,15}}, we have derived a set of
interesting informations on band structure of gravitinos and gauginos in
$\mathcal{N}=2$ theory. Part of these information have been obtained from the
proposal (\ref{13}); and its quantum extension that we rephrase here below:

$\left(  i\right)  $ the interpretation of the Andrianopoli realisation
$\vec{\xi}=\vec{\nu}\wedge \vec{m}$ as an angular momentum vector of a
quasi-particle with phase space coordinates $\left(  \vec{\nu},\vec{m}\right)
$ allowed us to think of the two gravitinos and the two gauginos in terms of
classical isospin $\frac{1}{2}$ particles (iso-particles) charged under
$U\left(  1\right)  _{elec}\times U\left(  1\right)  _{mag}$ gauge symmetry.
As a consequence of this observation, the scalar potential $\mathcal{V}$ has
been interpreted as the hamiltonian (\ref{ph}-\ref{hp}) of free iso-particle
and the central extension of the $\mathcal{N}=2$ supercurrent algebra
(\ref{ac}) as describing isospin-orbit coupling $\vec{\xi}.\mathcal{\vec{I}}$.
This isospin-orbit interaction is the homologue of the usual spin-orbit
coupling $\vec{L}.\vec{S}$ in electronic systems of condensed matter. The
proposal (\ref{13}) allowed us also to derive two discrete symmetries $T$ and
$TP$ capturing data on partial breaking of $\mathcal{N}=2$ supersymmetry; see
subsection 3.3 for details. Exact $\mathcal{N}=2$ lives at the fix point of
these symmetries. In summary, we can say that the classical properties of the
isoparticle is given by the $\mathcal{N}=2$ supersymmetric current algebra
(\ref{ca}).

$\left(  ii\right)  $ By using Nielson-Ninomiya theorem, we have developed the
study of the topological property of fermionic gapless states given by zeros
of the discriminant (\ref{ml}). The two bands of the rigid Ward operator $H$
are gapped except at isolated points in the phase space of the electric and
magnetic coupling constants where supersymmetry is partially broken and where
live a gapless chiral state with a chiral anomaly violating the
Nielson-Ninomiya theorem. From the study of the properties of $H$; it follows
that the gap energy is given by $E_{g}=2\left \vert \vec{\xi}\right \vert $ and
vanishes for $\left \vert \vec{\xi}\right \vert =0$; that is for vanishing
central extension in the $\mathcal{N}=2$ supercurrent algebra. Zero modes of
$H$ and their properties like windings and conserved supersymmetric charges
are as reported in table (\ref{t}).$\ $At the particular point $\mathcal{V}%
=0$, the discriminant $\det H$ reduces to $-\left \vert \vec{\xi}\right \vert
^{2}$ and has an SU$\left(  2\right)  $ singularity at the origin $\vec{\xi
}=\vec{0}$. There, the Nielson-Ninomiya theorem $\sum_{i}w\left(  S_{i}%
^{2}\right)  =0$ is trivially satisfied as shown on the table (\ref{t}) and
$\mathcal{N}=2$ supersymmetry is exact with compensating chiral anomalies.

$\left(  iii\right)  $ We have used the proposal (\ref{13}) to study effect of
quantum corrections induced by fluctuations of FI coupling constants (running
couplings). We have found that quantum effect in iso-space of FI couplings may
break supersymmetry completely except for the saturated bounds (\ref{sat})
where half of osccillating modes diseapper.

Finally, we would like to add that this approach might be helpful to explore
the picture in higher supergravities; in particular for $\mathcal{N}=4$;
progress in this direction will be reported in a future occasion.

\end{document}